\documentclass[prb,floats,twocolumn,floatfix]{revtex4-1}
\usepackage{color}
\usepackage{amsmath}
\usepackage{amsfonts}
\usepackage{graphicx}
\newcommand{\bk}{{\bf k}}

\newcommand{\br}{{\bf r}}

\newcommand{\YBCO}{$\mathrm{YBa_2Cu_3O_{6+x}}$}
\newcommand{\BSCCO}{$\mathrm{Bi_2Sr_2CaCu_2O_{8+x}}$}

\begin{document}
\title{Intrinsic Mechanism for Magneto-Thermal  Conductivity Oscillations in Spin-Orbit-Coupled Nodal Superconductors}
\date{\today}
\author{W. A. Atkinson}
\affiliation{Department of Physics and Astronomy, Trent University, Peterborough, Ontario K9L 0G2, Canada}
\email{billatkinson@trentu.ca}
\author{A. P. Kampf}
\affiliation{Theoretical Physics III, Center for Electronic Correlations and Magnetism, Institute of Physics, University of Augsburg, 86135 Augsburg, Germany}
\begin{abstract}
{
We describe a mechanism by which the longitudinal thermal conductivity $\kappa_{xx}$, measured in an in-plane magnetic field, oscillates as a function of field angle in layered nodal superconductors.  These oscillations occur when the spin-orbit splitting at the nodes is larger than the nodal scattering rate, and are complementary to vortex-induced oscillations identified previously.  In sufficiently anisotropic materials, the spin-orbit mechanism may be dominant. As a particular application, we focus on the cuprate high-temperature superconductor YBa$_2$Cu$_3$O$_{6+x}$.  This material belongs to the class of Rashba bilayers, in which individual CuO$_2$ layers lack inversion symmetry although the crystal itself is globally centrosymmetric.  We show that spin-orbit coupling endows $\kappa_{xx}/T$ with a characteristic dependence on magnetic field angle that should be easily detected experimentally, and argue that for underdoped samples the spin-orbit contribution is larger than the vortex contribution.  A key advantage of the magneto-thermal conductivity is that it is a bulk probe of spin-orbit physics, and therefore not sensitive to inversion breaking at surfaces.}
\end{abstract}
\maketitle

\section{Introduction}

Nodal superconductors are characterized by an energy gap that vanishes at point or line ``nodes'' on the Fermi surface.  Low-energy quasiparticle excitations exist in the neighborhood of the nodes, and these excitations are reflected in characteristic power laws in the temperature dependence of various thermodynamic quantities.\cite{matsuda_nodal_2006}  These power laws can distinguish different node types (i.e.\ line versus point), but contain incomplete information about the symmetry of the superconducting state.  Magneto-thermal conductivity measurements are particularly useful in this regard, as the magnetic field dependence contains  information about the $k$-space structure of the gap nodes.\cite{matsuda_nodal_2006}

A typical experiment involves the measurement of the thermal conductivity tensor $\kappa_{ij}$ in a magnetic field, which is swept through polar and/or azimuthal angles.  In a nodal superconductor, the longitudinal thermal conductivity, for example $\kappa_{xx}$, will oscillate as a function of the relative orientation of the field and the gap nodes.\cite{vekhter_anisotropic_1999,vorontsov_unconventional_2007}  The oscillation pattern has a nontrivial dependence on field strength and temperature, but with appropriate modeling will reveal the symmetry of the superconducting state.\cite{vorontsov_unconventional_2007,das_field-angle-resolved_2013}  Experimentally, this technique has been used to study the gap symmetry for  a variety of unconventional superconductors, including organic\cite{izawa_superconducting_2002} and heavy Fermion materials,\cite{watanabe_superconducting_2004,matsuda_nodal_2006,kim_resonances_2017} optimally doped YBa$_2$Cu$_3$O$_{7-\delta}$,\cite{yu_tensor_1995,aubin_angular_1997,ocana_thermal_2002}  and Sr$_2$RuO$_4$.\cite{izawa_superconducting_2001}

There are two established mechanisms underlying these oscillations, both of which are associated with the vortex lattice formed by the magnetic field.  At low temperatures, circulating vortex currents ``Doppler shift'' the quasiparticle energies by an amount $\hbar {\bf v}_s(\br)\cdot \bk$, where $\bk$ is the quasiparticle wavevector and ${\bf v}_s(\br)$ the superfluid velocity in the neighborhood of $\br$.\cite{volovik_superconductivity_1993,vekhter_anisotropic_1999}  The Doppler shift induces a nonzero density of states at each node that depends on the angle between the Fermi wavevector $\bk_F$ at that node and ${\bf v}_s(\br)$.  The total induced density of states, and consequently the thermal conductivity, therefore changes with the orientation of the vortex lattice or, equivalently, the field angle.  At high temperatures, a second mechanism takes over, namely anisotropic quasiparticle scattering by vortices becomes the dominant source of field-angle dependence.\cite{vorontsov_unconventional_2007}

In this work, we discuss a third mechanism that is especially relevant to layered superconductors in the highly anisotropic (quasi-two-dimensional) limit. In this geometry, an in-plane magnetic field generates only weak circulating vortex currents because of the small quasiparticle mobility along the interlayer direction.  Under such circumstances, we show that one may still observe pronounced field-angle oscillations in the presence of spin-orbit coupling (SOC).  As a particular application of this mechanism, we focus on so-called ``hidden spin-orbit'' superconductors.

In the past few years, a number of materials have been discovered that exhibit signatures of SOC despite being both centrosymmetric and time-reversal symmetric.\cite{Riley:2014,Santos-Cottin:2016,Gehlmann:2016,Razzoli:2017,Wu:2017,Yao:2017} Naive considerations would suggest this is not possible:  by Kramers' theorem, materials that satisfy both inversion and time-reversal symmetry must have degenerate energies $E_{\bk \uparrow}$ and $E_{\bk \downarrow}$, which seems to eliminate the possibility of $k$-space spin textures.  Such spin textures, which are a hallmark of SOC, have nonetheless been observed.   Key to this is that the Kramers-degenerate states are spatially separated, which leads to spin textures that are localized in space.\cite{Zhang:2014jw,Liu:2015,Yuan:2019}  
    
 Rashba bilayers form a prominent subclass of hidden spin-orbit materials. In these materials, the unit cell contains pairs of conducting layers; while the unit cells are centrosymmetric, the individual layers are not.  Rather, the layers are ``inversion pairs'', meaning that they transform into one another under inversion.\cite{Liu:2015,Yuan:2019}  The individual layers thus exhibit some combination of Rashba and Dresselhaus SOC, with the Rashba contribution typically being larger in layered materials;\cite{Liu:2015} however, global inversion symmetry guarantees that spin textures in one layer are compensated for by opposite textures in the other layer.  It is thus essential that the coupling between the layers be weak, as the $k$-space spin textures will be quenched when the two layers are strongly hybridized.
      
 Rashba bilayers have been investigated as possible topological insulators,\cite{Das:wl} as semimetals with electrically tunable Dirac cones,\cite{Dong:2015vo} and as model systems with nontrivial superconducting\cite{Sigrist:2014,Higashi:2016,Yoshida:2012df,Liu:2017} and nematic\cite{Hitomi:2014,Hitomi:2016} phases. Furthermore, many high-temperature superconductors, including \YBCO{} (YBCO$_{6+x}$) and \BSCCO{} (Bi2212), satisfy the structural requirements to be Rashba bilayers; however, the relevance of this fact to cuprate physics is not established and hinges on the size of the effect.
     
 Rashba-like spin polarization patterns have been directly measured in Bi2212 via spin-polarized angle-resolved photoemission spectroscopy (ARPES) experiments.\cite{Gotlieb:2018fx}  While important, these observations require independent confirmation because ARPES is a surface probe, and therefore sensitive to inversion symmetry breaking at surfaces.  Indirect evidence for SOC has been obtained from the magnetic breakdown energy scale that one infers from many quantum oscillation experiments in YBCO$_{6+x}$.\cite{Audouard:2009,Sebastian:2012}   However, it remains open whether the observed splitting is due to spin-orbit physics,\cite{Harrison:2015jj,Briffa:2016fs} or to interlayer coupling.\cite{Maharaj:2016}  A recent microscopic model for YBCO$_{6+x}$\cite{Atkinson:2020} suggests that the Fermi surface is spin-split by $\sim 10$--20~meV due to a Rashba-like SOC; however, this is an upper bound as interlayer coupling may quench spin-orbit physics.
     
 In this work, we show that the nodal structure of the $d$-wave superconducting gap allows for an elegant and straightforward observation of spin-orbit coupling through the longitudinal thermal conductivity in a transverse magnetic field.  This effect will be present whether or not the superconductor is quasi-two-dimensional (quasi-2D), but must be disentangled from Doppler shift contributions if circulating vortex currents are not negligible. Importantly, this is a bulk measurement that is insensitive to inversion symmetry breaking at sample surfaces, and is complementary to other recent proposals:  Kaladzhyan {\em et al.} showed that the dominant Friedel oscillation wavevectors associated with impurity scattering (as measured by scanning tunneling spectroscopy) reflect the spin-splitting of the Fermi surface and can be used to obtain the spin-orbit coupling constant,\cite{Kaladzhyan:2016}  while Raines {\em et al.} discussed the practicality of spin-Hall and Edelstein effects as probes of SOC.\cite{Raines:2019}

We provide a simple description of the effect in Sec.~\ref{sec:overview}.  While the mechanism has some similarities to the Doppler shift mechanism described in Ref.~\onlinecite{vekhter_anisotropic_1999}, there is the essential difference that a finite density of states is induced by Zeeman coupling to the quasiparticles rather than by circulating currents.  Thermal conductivity calculations are described in Sec.~\ref{sec:calcs}, although details are left to the appendices, and results of these calculations are given in Sec.~\ref{sec:results}.  We address the important question of how to distinguish spin-orbit and Doppler-shift contributions to the magneto-thermal conductivity in Sec.~\ref{sec:discussion}.  In that same section, we make an estimate that suggests that magneto-thermal oscillations in  YBa$_2$Cu$_3$O$_{6.5}$ are dominated by spin-orbit effects.
     
\section{Origin of the Density of States Oscillations}
\label{sec:overview}     
     
\begin{figure*}
     \includegraphics[width=\textwidth]{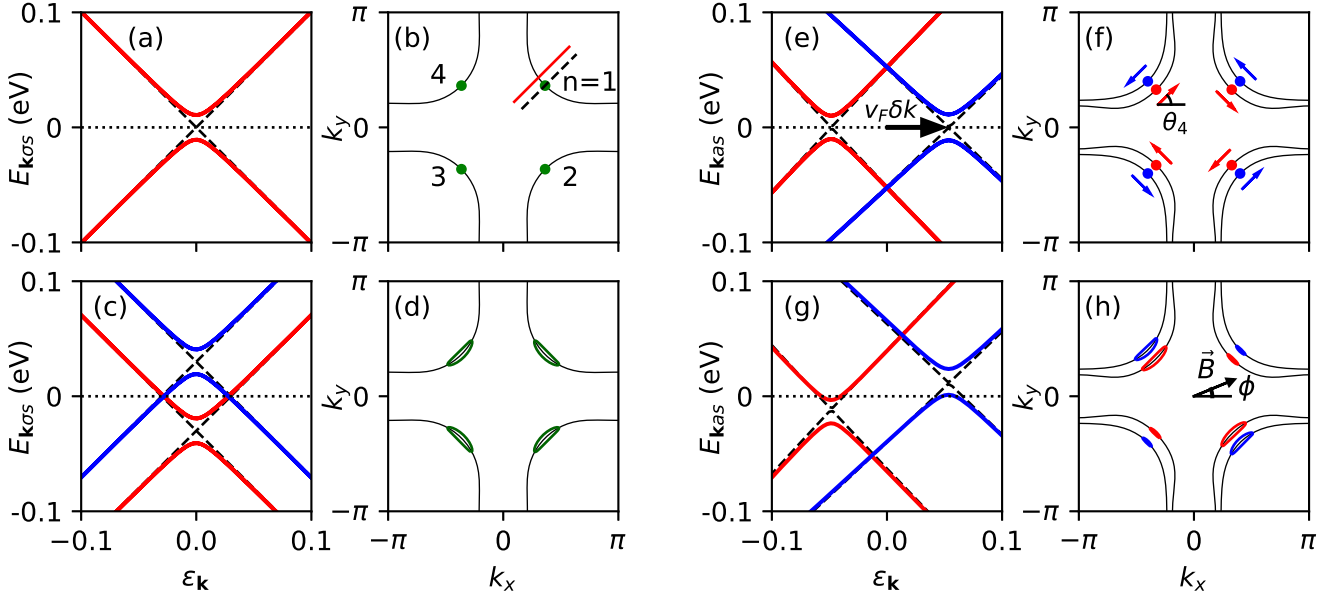}
     \caption{Effects of a Zeeman field on quasiparticle dispersion for a 2D $d_{x^2-y^2}$ superconductor.  Results are shown (a-d) for vanishing SOC and (e-h) for a nonzero Rashba SOC. (a) Without SOC, there are two doubly-degenerate quasiparticle branches, with energies $E_{\bk \sigma s} = s E_\bk$, with $\sigma$ the spin index and $s=\pm$ the band index. The solid (dashed) branches are plotted along the solid (dashed) cut through the Brillouin zone indicated in (b).   The dispersion is gapless along cuts through any nodal point (dashed line), but is otherwise gapped (solid line). (b) For a $d_{x^2-y^2}$ superconductor, there are four nodal points (green dots) located on the normal-state Fermi surface (black line) and labeled by a nodal index $n$.    (c) A magnetic Zeeman field rigidly shifts the quasiparticle branches by energies $\pm\mu B$, such that the branches are no longer degenerate. (d) These shifts inflate the nodal points to form Bogoliubov Fermi surfaces (green ellipses). (e) With SOC, bands are labeled by their helicity $a=\pm$ and band index $s=\pm$.  SOC shifts the gap nodes shown in (b) by $-a \delta k$, with the red (blue) bands corresponding to $a=+$ ($a=-$).  (f)  Spin-momentum locking on the normal-state Fermi surface determines the spin polarization of quasiparticle states near the gap nodes.  Arrows correspond to the spin polarizations of the bands in (e).  For node $n$, $\theta_{n}$ is defined as the smallest angle between the nodal spin polarization and the horizontal axis for the positive helicity band.   (g) In an in-plane magnetic Zeeman field, the quasiparticle branches are shifted by $-a\mu B \cos(\theta_{n}+\phi)$, where $\phi$ is the angle between the field and the horizontal axis.  (h)  As a result, the sizes of the Bogoliubov Fermi surfaces are node-dependent.  Note that the SOC ($\alpha = 40$~meV) and the field strength ($\mu B = 30$~meV) are artifically inflated for clarity.  Other parameters are as described in the main text.}
         \label{fig:overview}
\end{figure*}
     
{
	
For a singlet superconductor, the BCS Hamiltonian takes the form
\begin{eqnarray}
\hat H_\mathrm{BCS} &=& \sum_{\bk,\sigma} \epsilon_\bk c^\dagger_{\bk\sigma}c_{\bk\sigma}  \nonumber \\
&& + \frac 12 \sum_\bk \left[ 
\chi_\bk \left( c^\dagger_{\bk\uparrow}c^\dagger_{-\bk\downarrow} - c^\dagger_{\bk\downarrow}c^\dagger_{-\bk\uparrow} \right ) + \mathrm{h.c.}\right ]
\label{eq:HBCS}
\end{eqnarray}	
where $\mathrm{h.c.}$ is the hermitian conjugate, $\epsilon_\bk$ is the normal-state dispersion measured relative to the chemical potential, and $\chi_\bk$ is the superconducting order parameter.  The pairing term is typically simplified by making the permutation $-c^\dagger_{\bk \downarrow} c^\dagger_{-\bk\uparrow}= c^\dagger_{-\bk \uparrow} c^\dagger_{\bk\downarrow}$ and recognizing that $\chi_\bk = \chi_{-\bk}$.  However, Eq.~(\ref{eq:HBCS}) can be  extended easily to include SOC, and is therefore left as-is for this discussion.
	
Written in this form, Eq.~(\ref{eq:HBCS}) generates four flavors of BCS quasiparticle, described by the creation operators
\begin{eqnarray}
\gamma_{\bk \uparrow +}^\dagger &=& u_\bk c^\dagger_{\bk \uparrow} + v^\ast_\bk c_{-\bk \downarrow},
\label{eq:g1p} \\
\gamma_{\bk \uparrow -}^\dagger &=&  -v_\bk c^\dagger_{\bk \uparrow} + u_\bk c_{-\bk \downarrow},
\label{eq:g1m} \\
\gamma_{\bk \downarrow +}^\dagger &=& u_\bk c^\dagger_{\bk \downarrow} - v^\ast_\bk c_{-\bk \uparrow},
\label{eq:g2p} \\
\gamma_{\bk \downarrow-}^\dagger &=&  v_\bk c^\dagger_{\bk \downarrow} + u_\bk c_{-\bk \uparrow},
\label{eq:g2m}
\end{eqnarray}
where the coherence factors are
\begin{eqnarray}
u_\bk = \frac{1}{\sqrt{2}} \sqrt{1+\frac{\epsilon_\bk}{E_\bk}},  \quad
v_\bk = \frac{\chi_\bk}{\sqrt{2}|\chi_\bk|} \sqrt{1-\frac{\epsilon_\bk}{E_\bk}}, 
\end{eqnarray}
and $E_\bk = \sqrt{\epsilon_\bk^2 + \chi_{\bk}^2}$ is the usual BCS quasiparticle excitation energy. The operators $\gamma^\dagger_{\bk \sigma s}$ defined by  Eqs.~(\ref{eq:g1p})--(\ref{eq:g2m}) are labeled by their spin-state $\sigma$ and band index $s = \pm$;  we have followed the convention that the quasiparticle spectrum has two branches, with energies $E_{\bk \sigma \pm} = \pm E_\bk$, corresponding to the quasiparticle operators $\gamma_{\bk \sigma \pm}^\dagger$.  The branches are independent of $\sigma$, and are thus doubly degenerate with the upper (lower) branches empty (fully occupied) at zero temperature.

These branches are pictured in Fig.~\ref{fig:overview}(a) along two cuts through the Brillouin zone for the case of a nodal $d_{x^2-y^2}$ superconductor.  The locations of the cuts  are indicated in Fig.~\ref{fig:overview}(b), which shows also the normal-state Fermi surface and gap nodes, i.e.\ points on the Fermi surface where $\chi_{\bk}$ vanishes such that $E_\bk=0$.  In Fig.~\ref{fig:overview}(a), the excitation branches disperse linearly near the node, and are gapped away from the node. 

The quasiparticles defined by Eqs.~(\ref{eq:g1p})--(\ref{eq:g2m}) have  well-defined spins, such that the z-component of the spin operator is
\begin{eqnarray}
\hat S_z &=& \frac 12 \sum_\bk (c^\dagger_{\bk\uparrow} c_{\bk\uparrow} - c^\dagger_{\bk\downarrow} c_{\bk\downarrow}), \nonumber \\
&=& \frac 14 \sum_{\bk}\sum_{s=\pm} (\gamma^\dagger_{\bk \uparrow s } \gamma_{\bk \uparrow s} - \gamma^\dagger_{\bk \downarrow s} \gamma _{\bk \downarrow s}).
\end{eqnarray}
A magnetic Zeeman field adds a term $-g \mu_B B \hat S_z$ to the electronic Hamiltonian, with $g$ the electronic g-factor and $\mu_B$ the Bohr magneton.
Because $\hat S_z$ is diagonal in the quasiparticle operators $\gamma_{\bk \sigma s}$, this additional term leaves the quasiparticles intact, but rigidly shifts their dispersions by an amount $-\mu B$ ($\sigma=\uparrow$) or $+\mu B$ ($\sigma=\downarrow$), where $\mu = \frac 12 g \mu_B$ is the electron dipole moment.  These shifts are independent of the field direction, as the spin quantization axis is arbitrary in the absence of SOC, and is the same for each of the nodes.   The resultant superconducting bands are shown in Fig.~\ref{fig:overview}(c) along the same cuts as in Fig.~\ref{fig:overview}(a). The band shifts inflate the nodal points to form so-called Bogoliubov Fermi surfaces\cite{agterberg_bogoliubov_2017} that separate occupied and empty quasiparticle states.  These are shown as green ellipses in Fig.~\ref{fig:overview}(d); they are the same for all nodes and are independent of field direction.\cite{Yang:1998} 

The situation changes when the SOC is nonzero.  Here, the size of the induced Bogoliubov Fermi surfaces varies from node to node and depends on the field angle.  
For a generic 2D $d_{x^2-y^2}$ superconductor with Rashba SOC, the Hamiltonian is 
$\hat H = {\sum_{\bk}}^\prime {\bf C}_\bk^\dagger {\bf H}_\bk  {\bf C}_\bk$, with 
\begin{eqnarray}
{\bf H}_\bk = \left [ \begin{array}{cc}
{\bf h}_\bk & \Delta_\bk \\
\Delta^\dagger_\bk & - {\bf h}^T_{-\bk}
\end{array} \right ],
\label{eq:Hsoc}
\end{eqnarray} 
where ${\bf C}_\bk = (\begin{array}{cccc} 
c_{\bk \uparrow}, & c_{\bk \downarrow}, & c^\dagger_{-\bk \uparrow}, & c^\dagger_{-\bk \downarrow} \end{array}  )^T$, the prime indicates the summation is over a reduced Brillouin zone $(k_x,k_y) \in [0,\pi]\otimes [-\pi,\pi]$, and
\begin{eqnarray}
{\bf h}_\bk &=& \epsilon_{\bk} \tau_0 + ( {\bf g}_\bk - \mu {\bf B} ) \cdot {\boldsymbol \tau} \label{eq:hk_soc} \\
{\bf \Delta}_\bk &=& i\tau_y \chi_{d\bk} \label{eq:Dk} \\
{\bf g}_\bk &=& \alpha (\sin k_y, -\sin k_x,0).
\end{eqnarray}
Here, ${\bf h}_\bk$ is the Hamiltonian for the normal state,   
${\boldsymbol\tau}$ and $\tau_0$  are the Pauli spin matrices, ${\bf g}_\bk$ is the Rashba spin-orbit term, and ${\bf B}$ is the in-plane magnetic field.    As shown in Fig.~\ref{fig:overview}(f), SOC splits the normal-state Fermi surface  into spin-polarized bands, with the electron spins locked to their momentum.

For the $d_{x^2-y^2}$ superconductor, the nodes shown in Fig.~\ref{fig:overview}(a) are shifted by displacements $\pm \delta k$ [Figs.~\ref{fig:overview}(e) and (f)], where $\delta k \sim \alpha/v_F$.  This doubles the number of nodes in each quadrant of the Brillouin zone, but the dispersion near each of the shifted nodes has the same structure  as when SOC is absent [Fig.~\ref{fig:overview}(e)]. (Note that although calculations are performed in a reduced Brillouin zone, we continue to show the full zone for illustrative purposes.)
We remark that the superconducting order parameter ${\bf \Delta}_\bk$ must develop a triplet component in response to the SOC, and that this may alter the structure of the gap nodes.  For simplicity, we make the assumption that the triplet component is small and can be neglected, which is certainly the case in YBCO$_{6+x}$ (see Appendix \ref{app:SC}).

In most SOC materials, the coupling constant $\alpha$ is orders of magnitude larger than $\mu B\sim 1$~meV, and in this limit the physics of the nodal dispersion is easily understood.  Crucially, the SOC selects a preferred polarization axis near each of the gap nodes, and this axis is largely unchanged by the Zeeman field in the limit $\mu B \ll \alpha$.  In each quadrant (labeled $n=1,\ldots,4$) one may locally rotate the spin-quantization axis such that the Hamiltonian for each band has a BCS-like form (Appendix \ref{sec:analytic}).
The quasiparticle creation operators are then similar to Eqs.~(\ref{eq:g1p})--(\ref{eq:g2m}) near the gap nodes, but with ``up'' and ``down'' spin directions aligned with the red and blue arrows, respectively, in Fig.~\ref{fig:overview}(f).  For quadrant $n$, we denote the angle between the ``up'' direction and the $k_x$ axis by $\theta_{n}$.  

In the limit $\mu B \ll \alpha$, the principal effect of the Zeeman field is to shift the nodal dispersions by an amount $-a\mu B \cos(\theta_{n}+\phi)$, where $\phi$ is the angle between the magnetic field and the $k_x$ axis, and $a=\pm$ is the helicity of the band (positive helicity indicates that the spin winds clockwise around the center of the Brillouin zone).  This has several consequences.  First, the dispersions near the two spin-split nodes are shifted in opposite directions because their helicities are opposite [Fig.~\ref{fig:overview}(g)]; however, the size of the induced nodal Fermi surfaces is nearly the same [Fig.~\ref{fig:overview}(h)] and the spin-split nodes make nearly identical contributions to the thermal conductivity.  Second, nodes belonging to different quadrants of the Brillouin zone  experience different shifts reflecting the different values of $\theta_{n}$, so that the sizes of the induced Bogoliubov Fermi surfaces are different [Fig.~\ref{fig:overview}(h)].  Finally, the relative sizes of the different nodal Fermi surfaces depend on the angle $\phi$ of the magnetic field.

}

\begin{figure}
	\includegraphics[width=\columnwidth]{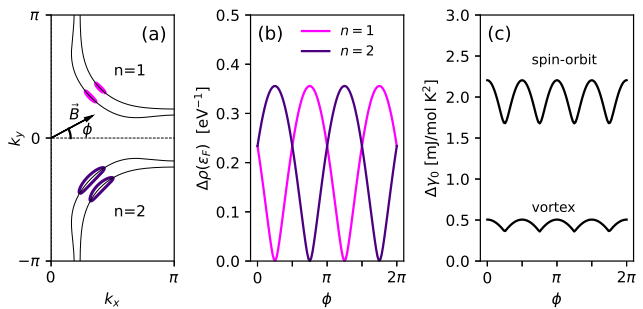}
	\caption{ Density of states induced by an in-plane Zeeman field as a function of field angle.	
	(a) Fermi surface structure in the reduced Brillouin zone.  The sizes of the Bogoliubov Fermi surfaces depend on the nodal index $n$ and on the field angle $\phi$. Fermi surfaces with $n=1$ (magenta) vanish when $\phi=(m+\frac 14)\pi$, while those with $n=2$ (violet) vanish when $\phi=(m-\frac 14)\pi$, with $m\in \mathbb{Z}$.  (b)  The density of states (per unit cell) induced by the Zeeman field is shown for each pair of gap nodes in (a). $\Delta \rho(\varepsilon_F)$ is calculated using Eq.~(\ref{eq:drho}) with realistic parameters for YBa$_2$Cu$_3$O$_{6.5}$   ($a_0 = 3.8$~\AA, $v_F = 1.2$~eV\AA, $v_2 = 0.017$~eV\AA, and $\gamma = 0.1$~meV).  (c) Corresponding oscillations of the  specific heat coefficient $\Delta \gamma_0$ on field angle are a signature of Rashba spin-orbit coupling.  For comparison, the estimated specific heat coefficients from circulating vortex currents are included (see Sec.~\ref{sec:discussion}).  Results in (b) and (c) are for $T\rightarrow 0$, and $\mu B = 1$~meV, which corresponds to $B=17$~T. }
	\label{fig:overview2}
\end{figure}

The sizes of the Fermi surface pockets [Fig.~\ref{fig:overview2}(a)] are directly related to the induced density of states  $\Delta \rho(\varepsilon_F)$ at the Fermi energy.   Analytic expressions for $\Delta \rho(\varepsilon_F)$ may be obtained  (Appendix~\ref{sec:DOS}), and their field-angle-dependence, 
shown for each of the nodal regions in Fig.~\ref{fig:overview2}(b) is a signature of Rashba SOC.  The two regions oscillate out of phase with each other, and the total induced density of states obtained from their sum has minima at field angles $\phi=(m\pm \frac 14)\pi$.  This is reflected in the linear specific heat coefficient,
\begin{equation}
\Delta \gamma_0 = \lim_{T\rightarrow 0} \frac {\Delta c_v}{T} = \frac{\pi^2}{3} k_B^2 \Delta \rho(\epsilon_F),
\end{equation}
shown in Fig.~\ref{fig:overview2}(c).
The size of the oscillations depends on both ${\bf B}$ and on the single-particle scattering rate $\gamma$. For the benchmark case of YBa$_2$Cu$_3$O$_{6.5}$ shown in Fig.~\ref{fig:overview2}(c), the induced specific heat coefficient $\Delta \gamma$ is comparable to typical measured values for an out-of-plane magnetic field (i.e.\ for the vortex phase).\cite{Moler:1997} Significantly, the predicted magneto-thermal oscillations are a factor of 4 larger than the expected vortex contributions for an {\em in-plane} field.  In Fig.~\ref{fig:overview2}(c), the vortex contribution is calculated using an estimate from Ref.~\onlinecite{vekhter_anisotropic_1999} and is discussed in detail in Sec.~\ref{sec:discussion}.
  
To obtain quantitative estimates for YBa$_2$Cu$_3$O$_{6.5}$, we have used a dispersion $\epsilon_\bk$ that was obtained from a tight-binding fit to ARPES measurements on YBCO$_{6+x}$,\cite{Pasanai:2010ky} and a realistic gap function $\chi_{d\bk} = \chi_d (\cos k_x - \cos k_y)/\sqrt{2}$ with  $\chi_{d} = 50$~meV.  These choices give a nodal Fermi velocity $v_F = 1.2$~eV\AA, in close agreement with experiments,\cite{Vishik:2010}  and a superconducting nodal group velocity $v_2 = |\nabla_\bk \Delta_\bk|_\mathrm{node} = 0.17$~eV\AA.  Unless specified otherwise, the Rashba coupling constant is taken to be $\alpha=10$~meV throughout this work, which gives a spin-splitting at the gap nodes of 25~meV.\cite{Atkinson:2020} 

Although our explanation of the density of states oscillations is based on the weak-field limit, the effect is general provided the spin-splitting of the nodes is greater than the quasiparticle scattering rate $\gamma$.   The effect is present regardless of the dimensionality of the system and will dominate over the vortex contribution in highly anisotropic materials; however, even in three dimensional materials the SOC and vortex  effects can be comparable.


We finish this section with a comment that is specific to Rashba bilayers.
The model explored in this section describes a single CuO$_2$ layer, and there are two issues that might limit its applicability to the bilayer.  First, in a Rashba bilayer, the sign of $\alpha$ is opposite in each layer, and it is a concern that the contributions from each layer might cancel.  However, from Fig.~\ref{fig:overview2}(a), it is clear that reversing the direction of each nodal polarization will have no effect on the induced density of states.  Indeed, as we show explicitly below, the  field-angle dependence of the thermal conductivity is an even function of $\alpha$.  Second, one must keep in mind that hybridization of the two layers via a hopping matrix element $t_\perp$ will quench the spin polarization.  As discussed elsewhere,\cite{Harrison:2015jj,Briffa:2016fs,Atkinson:2020} the spin polarization at the gap nodes is of order $\alpha/\sqrt{t_\perp^2+\alpha^2}$.  The analysis contained in this work assumes that $t_\perp \ll \alpha$, which is supported by an apparent collapse of bilayer splitting at the gap nodes in underdoped YBCO$_{6+x}$.\cite{Fournier:2010kk} An experimental failure to measure the predicted thermal conductivity oscillations likely implies that the limit $t_\perp \ll \alpha$ does not apply.

\section{Thermal Conductivity}
\subsection{Calculations}
\label{sec:calcs}

In this section, we discuss calculations of the longitudinal thermal conductivity in the presence of an in-plane magnetic Zeeman field.  As in the previous section, we assume that the triplet contribution to the superconducting order parameter can be neglected.  For YBCO$_{6+x}$, we have checked numerically that neglect of the triplet components has no observable effect on the calculated longitudinal thermal conductivity. This is essentially different, then, from the intrinsic thermal Hall effect in a perpendicular Zeeman field, which depends crucially on the triplet component;  together with SOC, a perpendicular Zeeman field creates a gapful mixed-parity topological superconductor\cite{Yoshida:2016,Daido:2016,daido_majorana_2017} whose finite Chern number determines the $T$-linear part of the thermal Hall conductivity.\cite{Sumiyoshi:2013eu}

Following Refs.~\onlinecite{Ambegaokar:1965bz,Durst:2000iw}, the thermal current operator is
\begin{eqnarray}
{\bf J}_Q  &=&  -\frac i2 {\sum_{\bk}}' \sum_{i,j}
{\bf V}_{\bk, ij} \left (  \dot{C}^\dagger_{\bk, i} {C}_{\bk, j}
- {C}^\dagger_{\bk, i} \dot {C}_{\bk, j} \right )
\label{eq:JQ0}
\end{eqnarray}
where $\dot {C}_{\bk, i}$ indicates a time derivative of ${C}_{\bk,i}$, and
where the  velocity matrix is
\begin{equation}
{\bf V}_\bk = 
\left [ \begin{array}{cc} {\bf v}_\bk & {\bf v}_{\Delta,\bk} \\ 
{\bf v}_{\Delta,\bk}^\dagger & {\bf v}_{-\bk}^T  
\end{array}\right ].
\label{eq:VSC}
\end{equation}
In this expression, ${\bf v}_\bk = \nabla_\bk {\bf h}_\bk$ and ${\bf
  v}_{\Delta,\bk} = \nabla_\bk {\bf \Delta}_\bk$, with ${\bf h}_\bk$ and ${\bf \Delta}_\bk$ given by Eqs.~(\ref{eq:hk_soc}) and (\ref{eq:Dk}), respectively.  Equation~(\ref{eq:JQ0}) does not include corrections due to circulating thermal currents that appear when time-reversal symmetry is broken,\cite{Qin:2011fk} as these do not contribute to the longitudinal thermal conductivity.

From the Kubo formula, the longitudinal thermal conductivity satisfies
\begin{equation}
\frac{\kappa_{xx}}{T} = -\frac{\pi}{\hbar d T^2}   \int_{-\infty}^\infty dx \, x^2  \frac{ \partial f(x)}{\partial x} \Pi^{xx}(x). 
\label{eq:kxxT}
\end{equation}
where $f(x)$ is the Fermi function,  $d$ is the mean interlayer distance (for the bilayer case of YBCO$_{6+x}$, it is the $c$-axis lattice constant divided by two), and where 
\begin{equation}
\Pi^{xx}(x) = \frac{\hbar^2}{2N_k a_0^2} {\sum_{\bk}}' \mbox{Tr} \left [ {\bf A}_\bk(x) {\bf V}_{\bk} {\bf A}_\bk(x) {\bf V}_{\bk} \right ].
\label{eq:Pi}
\end{equation}
is the dimensionless thermal conductivity kernel.
In Eq.~(\ref{eq:Pi}),  $N_k$ is the number of k-points in the reduced Brillouin zone, $a_0$ is the lattice constant,  and ${\bf A}_\bk(x)$ is the spectral function obtained from the
Hamiltonian ${\bf H}_\bk$,
\begin{equation}
{\bf A}_\bk(x) = \frac{1}{2\pi i} \left [ ( x - i\gamma - {\bf H}_\bk )^{-1}
- ( x + i\gamma - {\bf H}_\bk )^{-1} \right],
\end{equation}
with $\gamma$ the quasiparticle scattering rate. 
At low $T$, Eq.~(\ref{eq:kxxT}) simplifies to
\begin{equation}
\frac{\kappa_{xx}}{T} = \frac{\pi^3 k_B^2}{3 \hbar d} \Pi^{xx}(0).
\label{eq:kxxT0}
\end{equation}

In the limit of strong SOC,  $|\alpha| \gg \mu B, \gamma$, it is further possible to obtain an analytic result for $\Pi^{xx}(x)$ (see Appendix \ref{sec:analytic}),
\begin{eqnarray}
\Pi^{xx}(x) &=& \frac{1}{8\pi^3}\frac {v_F^2+v_2^2}{ v_Fv_2} \Bigg \{ 
4 + \sum_\pm {\cal E}\left[ \frac{x \pm \mu B \cos(\frac \pi 4 + \phi)}{\gamma} \right ] 
\nonumber \\ 
&& +  \sum_\pm {\cal E}\left[ \frac{x \pm \mu B \cos( \frac{3\pi}{4} + \phi)}{\gamma} \right ] \Bigg \},  
\label{eq:Pi0}
\end{eqnarray}
with 
\begin{equation}
{\cal E}(y) = \left( y + \frac 1y \right ) \tan^{-1} y.
\end{equation}

\subsection{Results}
\label{sec:results}

\begin{figure}[tb]
\includegraphics[width=\columnwidth]{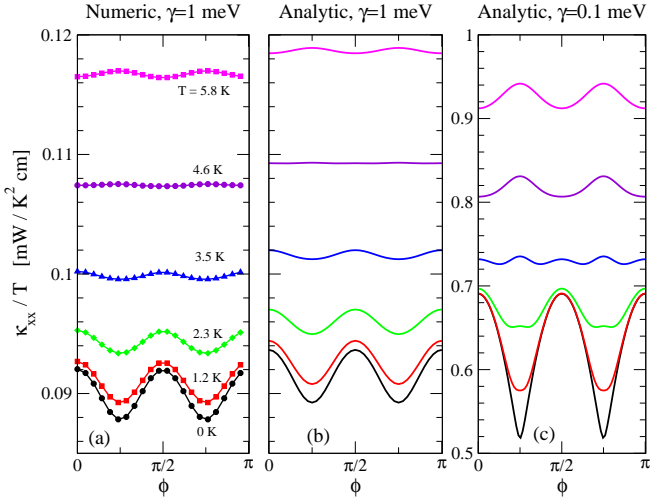} 
\caption{Thermal conductivity in the strong SOC limit ($|\alpha | \gg \mu B, \gamma$).  The thermal conductivity is  shown as a function of the angle $\phi$ between the magnetic field and the $x$-axis for a range of temperatures between $T=0$~K and $T = 5.8\mbox{ K} = 0.5$~meV.  Curves are generated for (a) the numerical evaluation of Eqs.~(\ref{eq:kxxT}) and (\ref{eq:Pi}), and (b, c) the analytical approximation Eq.~(\ref{eq:Pi0}).  The magnetic field is $\mu B = 2$~meV, which (assuming $\mu = \mu_B$) corresponds to $B=34$~T, and the SOC constant is  $\alpha = 10$~meV.  (a) and (b) are for an intermediate scattering rate, $\gamma = 1$~meV, and demonstrate that the analytical result is quantitatively accurate when the SOC is large; (c) is in the clean limit, $\gamma = 0.1$~meV, which is appropriate for YBCO$_{6.5}$.  Note that the scale in (c) is an order or magnitude larger than in (a) and (b).}
\label{fig:SLSC_Thermal}
\end{figure}

Figure~\ref{fig:SLSC_Thermal} shows the longitudinal thermal conductivity in an in-plane magnetic field, as a function of the angle between the field and the $x$-axis.  We focus initially on the large-SOC limit, with $\alpha = 10$~meV, since this is the regime that we expect to be relevant to YBCO$_{6+x}$.  
The essential point of this figure is that at sufficiently low temperatures, $\kappa_{xx}/T$ exhibits clear and pronounced oscillations.  The size of the oscillations is approximately proportional to $\mu B/\gamma$, and grows by an order of magnitude between the intermediate scattering [Fig.~\ref{fig:SLSC_Thermal}(a) and (b)] and clean [Fig.~\ref{fig:SLSC_Thermal}(c)] limits.  

The oscillations change their qualitative character as a function of temperature, and in the clean limit [Fig.~\ref{fig:SLSC_Thermal}(c)], the crossover between low- and high-temperature patterns occurs at $k_BT \sim \mu B$.  This crossover is shifted to somewhat higher temperatures when $\gamma\sim \mu B$ [Fig.~\ref{fig:SLSC_Thermal}(a) and (b)].  The shape of the oscillations depends also on the ratio $\mu B/\gamma$, taking an approximately sinusoidal form when $\mu B \sim \gamma$, and deviating strongly from it when $\mu B \gg \gamma$.

The figure also shows that the approximate expression (\ref{eq:Pi0}) for the transport kernel $\Pi^{xx}(x)$ works well when $\alpha$ is large.  Inspection of Eq.~(\ref{eq:Pi0}) reveals that, in the zero-temperature limit, $\kappa_{xx}/T$ is a function of $\mu B/\gamma$, and is independent of $\alpha$.  

\begin{figure}[tb]
	\includegraphics[width=\columnwidth]{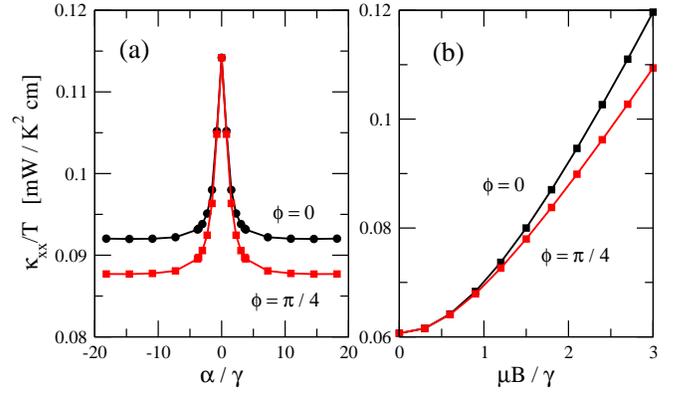} 
	\caption{Numerical results for the thermal conductivity as a function of (a) spin-orbit coupling strength and (b) magnetic field. Results are shown at field angles $\phi=0$ and $\phi=\pi/4$; in each plot, the difference between the two curves gives the amplitude of the oscillation.  Except where indicated otherwise, results are for $\alpha = 10$~meV, $\mu B = 2$~meV, $\gamma = 1$~meV, and $T=0$~K.  }
	\label{fig:alpha}
\end{figure}
This is apparent in Fig.~\ref{fig:alpha}(a), which shows that the oscillation amplitude at $T=0$ saturates at an approximately constant value when $|\alpha| \gg \gamma$.  We also note that  the amplitude is a symmetric function of $\alpha$, confirming our earlier assertion that the thermal conductivity oscillations due to the two layers making up a Rashba bilayer are additive.

Figure~\ref{fig:alpha}(b) shows the dependence of $\kappa_{xx}/T$ on magnetic field strength.  When $\mu B \rightarrow 0$, the conductivity kernel becomes independent of the scattering rate,\cite{Durst:2000iw}
\begin{equation}
\Pi^{xx}(0) \rightarrow \frac{1}{\pi^3} \frac{v_F^2+v_2^2}{v_Fv_2}, 
\end{equation}
yielding $\lim_{T\rightarrow 0} \kappa_{xx}/{T} = 0.062$~mW/K$^2\cdot$cm.  This value is close to that measured by Sutherland {\em et al.} in YBCO$_{6.54}$.\cite{Sutherland:2003de}
When $\mu B$ is not zero, the field dependence reflects the structure of the function ${\cal E}(\mu B/\gamma)$ in Eq.~(\ref{eq:Pi0}).  For small argument, the field dependence is quadratic, with ${\cal E}(\mu B/\gamma) \approx 1+ \frac 23 (\mu B/\gamma)^2$, while for large arguments it is linear with ${\cal E}(\mu B/\gamma) \approx \pi\mu B/2\gamma$.

The value of $\gamma$ is thus central to the observability of the thermal conductivity oscillations, which are suppressed when $\gamma > \mu B$.  Angle-resolved photoemission spectroscopy (ARPES) experiments have placed an upper bound of $\gamma =12$~meV on the nodal scattering rate for the bilayer cuprate superconductor Bi2212,\cite{Yamasaki:2007} consistent with the strong inhomogeneity observed in that material in tunneling experiments.\cite{pan_microscopic_2001,fischer_scanning_2007}  Spin-orbit effects might thus be hard to observe in Bi2212.  Conversely, microwave conductivity measurements  have found very small transport scattering rates $\gamma_\mathrm{tr} \lesssim 0.1$~meV in YBCO$_{6.50}$ at low temperatures.\cite{Harris:2006}  The reasonable assumption  $\gamma_\mathrm{tr} \sim \gamma$ places YBCO$_{6.50}$ in the clean limit, where thermal conductivity oscillations should be easily observable.

\begin{figure}[tb]
    \includegraphics[width=\columnwidth]{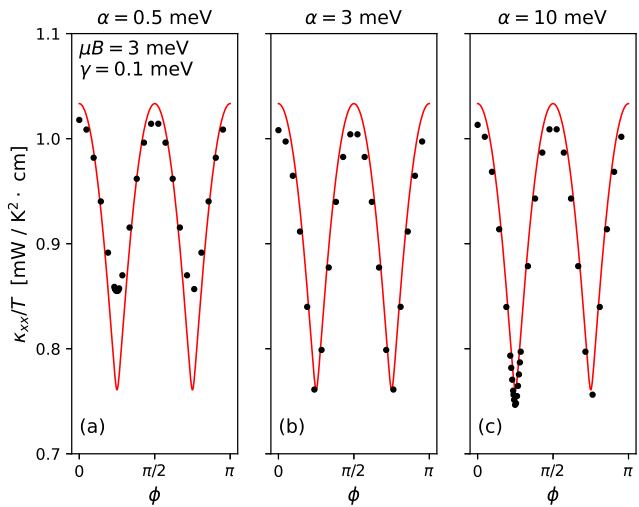} 
    \caption{Thermal conductivity in the clean limit, $\gamma \ll |\alpha|, \mu B$, as a function of field angle.  Numerical data (circles) are compared to the analytical result obtained from,  Eq.~(\ref{eq:kxxT0}) and  Eq.~(\ref{eq:Pi0}).  Equation~(\ref{eq:Pi0}) is nominally valid for $|\alpha| \gg \mu B$ but provides a good quantitative fit to the data even for $\alpha = \mu B$. Results are for $T=0$, $\mu B = 3$~meV, and $\gamma = 0.1$~meV. }
    \label{fig:thermal2}
\end{figure}


Figure~\ref{fig:thermal2} shows the range of behavior that can be expected for $\kappa_{xx}/T$ in the clean limit 
for different values of $\alpha$ spanning $|\alpha| \ll \mu B$ to $|\alpha| \gg \mu B$.  For comparison, the analytic result for large SOC is also plotted, and is quantitatively similar to the numerical data for all values of $\alpha$.  Remarkably, there is very little to distinguish the different thermal conductivity data sets, even though $\alpha$ changes by a factor of 20 across the figure.  Thus pronounced oscillations in the thermal conductivity should be observable so long as $|\alpha| > \gamma$.  By fitting Eq.~(\ref{eq:Pi0}) to experimental measurements of $\kappa_{xx}/T$ at low $T$ and for $\mu B \lesssim |\alpha|$, it is possible to extract both the ratio $v_F/v_2$ and the scattering rate $\gamma$.

\section{Discussion and Conclusions}
\label{sec:discussion}

In this work, we have shown that spin-orbit coupling generates a characteristic field-angle dependence of the longitudinal thermal conductivity in nodal superconductors.  These oscillations  reflect how the density of states induced by a magnetic Zeeman field depends on the angle between the field and the spin-polarization at the gap nodes.  Although we have focused on the case of a $d_{x^2-y^2}$ superconductor with Rashba SOC,  the mechanisms described in this work will be present for any nodal superconductor in which spin-orbit physics leads to spin-momentum locking at the gap nodes.  At low temperature ($T \lesssim \mu B$), the oscillation pattern is a function of the angle between the nodal spin axis and the magnetic field, and therefore depends both on the location of the nodes and the structure of the SOC.  When the SOC is known, the magneto-thermal oscillations can be used to determine the locations of the  nodes; conversely, when the node positions are known, the oscillations yield a fingerprint of the SOC.

This analysis is simplest if there is a clean separation between spin-orbit and vortex contributions to the magneto-thermal oscillations.  Indeed, the field-angle oscillations are qualitatively similar in both cases, and the question of how one may distinguish them is important.  For $d$-wave superconductors, it is common to write the field-angle dependence of the thermal conductivity as a series,
\begin{eqnarray}
\kappa_{xx} &=& \kappa_{0,xx} + \kappa_{2} \cos 2\phi + \kappa_{4} \cos 4\phi +\ldots
\label{eq:kxexpansion} \\
\kappa_{yy} &=& \kappa_{0,yy} - \kappa_{2} \cos 2\phi + \kappa_{4} \cos 4\phi +\ldots
\label{eq:kyexpansion}
\end{eqnarray}
The terms $\kappa_2$ and $\kappa_4$ are the amplitudes of the twofold- and fourfold-symmetric contributions to $\kappa_{xx}$, respectively, and $\phi$ is the field angle as before.  The fourfold term, $\kappa_4$, is a direct consequence of the fourfold symmetry of the excitation spectrum.  It is generally attributed to the symmetry of the gap function, but may also reflect the underlying band structure.\cite{das_field-angle-resolved_2013} 

$\kappa_4$ may be positive or negative, with $\kappa_{xx}$ having its minimum at $\phi = \pi/4$ in the first case, and its maximum at $\pi/4$ in the second.   For oscillations due to the vortex lattice, the general trend at low fields is that $\kappa_{xx}$ is positive at low $T$, but then goes negative as $T$ is increased.\cite{vorontsov_unconventional_2007}  Fig.~\ref{fig:SLSC_Thermal} shows that the trend is similar here, with the crossover happening at $k_BT \approx \mu B$ in the clean limit.  The $T$-dependence of $\kappa_4$ therefore does not allow us to isolate the source of the oscillations.
However, the field-strength dependence is qualitively different for the two mechanisms, and does allow one to distinguish between them.  In the clean limit ($\mu B \gg \gamma$) and at low $T$,  $\kappa_4$ grows as $\sqrt{B}$ for vortex-driven oscillations,\cite{vekhter_anisotropic_1999} but as $B$ for SOC-driven oscillations. 

Another distinction between SOC- and vortex-driven magneto-thermal conductivity oscillations lies in $\kappa_2$.  In previous work, $\kappa_2$ was found to arise from the scattering of thermal currents by the vortices, and the twofold anisotropy reflects the difference between driving currents parallel to and perpendicular to the vortices. While in many materials this term is dominant at elevated temperatures, it is suppressed in quasi-2D materials where the circulating currents are small.  Formally,  $\kappa_2 = 0$ in our calculations, and any nonzero value of $\kappa_2$ would therefore signal a nonzero vortex contribution.  A weak  $\kappa_2$ in conjunction with a significant $\kappa_4$ is a strong hint that SOC is the dominant factor in observed magneto-thermal conductivity oscillations.

One consequence of $\kappa_2$ vanishing is that, for the SOC-driven oscillations, the longitudinal thermal conductivity should be the same for both [100] and [110] heat currents provided the material is tetragonal,  and provided the transverse components $\kappa_{xy}$ and  $\kappa_{yx}$ vanish.  Under typical experimental conditions the heat current perpendicular to the applied temperature gradient is zero,   and the [100] and [110] longitudinal thermal conductivities are therefore\cite{vorontsov_unconventional_2007} 
\begin{equation}
\kappa_{l} = \left \{\begin{array}{ll}
\kappa_{xx} - \dfrac{\kappa_{xy}\kappa_{yx}}{\kappa_{yy}}, & [100]\\
 2\dfrac{\kappa_{xx}\kappa_{yy}-\kappa_{xy}\kappa_{yx}}{\kappa_{xx}+\kappa_{yy}+\kappa_{xy}+\kappa_{yx}} & [110]
\end{array} \right ..
\end{equation} 
%
For purely SOC-driven oscillations the intrinsic contribution to $\kappa_{xy}/T$ in the limit $T\rightarrow 0$ vanishes for a $d$-wave superconductor in an in-plane Zeeman field,\cite{Qin:2011fk,Sumiyoshi:2013eu} and the leading-order intrinsic contribution to $\kappa_{xy}$ must therefore be $O(T^2)$.  Since the leading-order contributions to $\kappa_{xx}$ and $\kappa_{yy}$ are $O(T)$, $\lim_{T\rightarrow 0} \kappa_l$ reduces to $\kappa_{xx}$ and $2\kappa_{xx}\kappa_{yy}/(\kappa_{xx}+\kappa_{yy})$ for the [100] and [110] directions, respectively. These are equal for tetragonal superconductors.

For the case of YBCO$_{6+x}$, we note that the material itself is orthorhombic due to the presence of one-dimensional CuO chains. This twofold anisotropy is tied to the crystal lattice, rather than the vortex lattice, and will show up primarily as a difference in the zeroth order term in the expansions of $\kappa_{xx}$ and $\kappa_{yy}$, so that $\kappa_{0,xx} \neq \kappa_{0,yy}$. 


We can estimate the relative importance of the vortex and SOC contributions for YBCO$_{6.5}$ by focusing on the specific heat.  At this doping level, YBCO$_{6+x}$ is strongly anisotropic, with the $c$-axis conductivity a factor of $10^3$ smaller than the in-plane conductivity. The density of states induced by circulating vortex currents can be obtained from the clean-limit ($\gamma\rightarrow 0$) approximation given in Ref.~\onlinecite{vekhter_anisotropic_1999}, 
\begin{equation}
\rho(\phi) \approx \rho_0 \frac{2\sqrt{2} E_H}{\pi \hbar v_2} \max \left( |\sin \phi|, |\cos \phi| \right ),
\end{equation}
with $\rho_0$ the normal state density of states and 
\begin{equation}
E_H \sim \frac{v_F}{2} \sqrt{\frac{\lambda_{ab}}{\lambda_c}} \sqrt{\frac{\pi B}{\Phi_0}},
\end{equation}
where $\Phi_0 = \pi \hbar/e$ is the superconducting flux quantum.  In YBCO$_{6.5}$, the ratio of in-plane and out-of-plane penetration depths $\lambda_c / \lambda_{ab} \approx 35$,\cite{bonn_microwave_1995,dulic_effects_2001} and the fitted dispersion $\epsilon_\bk$ gives the normal state density of states at the Fermi level $\rho_0 = 3.35$~eV$^{-1}$.  The resultant shift in the specific heat constant, $\Delta \gamma_0$ is shown in Fig.~\ref{fig:overview}(d) and is considerably smaller than that obtained from SOC.  It is reasonable to expect a similar disparity for the thermal conductivity, so that experimental observations of magneto-thermal conductivity oscillations for in-plane fields, along with a small value for $\kappa_2$, would be consistent with significant spin-orbit coupling.

We finish with a few caveats.  First, we note that a detailed quantitative description of any particular material will depend on the details of the scattering rate.  In particular, we have taken a simple model in which $\gamma$ depends on neither wavevector or energy.  While such an assumption is sensible for many materials, it can be problematic in nodal superconductors where the scattering rate can have a nontrivial energy dependence that is determined by the strength of the scattering potential.  In cuprates, for example, quantitatively accurate models typically require an admixture of Born and unitary scatterers.\cite{zhu_power_2004,lee-hone_disorder_2017,lee-hone_optical_2018}  Each of these has a characteristic energy dependence that will modify the temperature dependence of the oscillations.  Rigorous modeling of experiments will thus require a realistic disorder model.

Second, the question of how to extract the size of the spin-orbit coupling constant $\alpha$ remains open. While the existence of SOC in a nodal superconductor is easy to establish via the thermal conductivity oscillations, the amplitude of the spin-orbit coupling constant is harder to determine.  In clean materials, the size of the oscillations is nearly independent of $\alpha$ [Fig.~\ref{fig:alpha}(a)] and there is no clear crossover in behavior as a function of magnetic field strength between $\mu B \ll \alpha$ and $\mu B \gg \alpha$.  Rather, the amplitude of the oscillations is an indication of whether the spin splitting of the Fermi surfaces is large or small relative to the scattering rate.

Third, we have neglected in this work variations of the chemical potential with the field angle, which is another potential contribution to the field-angle dependence of $\kappa_{xx}$.  In our analytical calculations, these variations would manifest themselves as modulations of both the Fermi velocity and the anomalous velocity, $v_2$.  We have checked numerically that, at least for the model used in this work, chemical potential modulations have a negligible effect on the thermal conductivity.

In summary, we have demonstrated the existence of a novel intrinsic mechanism for oscillations of the longitudinal magneto-thermal conductivity as a function of magnetic field angle.  Zeeman coupling to nodal quasiparticles inflates the nodes into Bogoliubov Fermi surfaces, whose sizes depend on the field angle, the structure of the SOC, and the location of the gap nodes. The magnitude of the induced specific heat and thermal conductivity both depend on the sizes of the induced  Fermi surfaces, and the angle-dependence of the thermal properties provides a tool to explore the SOC (if the locations of the gap nodes are known) or the nodal structure (if the SOC is known).  As a probe of SOC, this technique has the advantage of being a bulk measurement, and is therefore insensitive to inversion-symmetry breaking at surfaces.

The structure of the magneto-thermal oscillations is qualitatively similar to what was found earlier for vortex-induced oscillations, but can be distinguished by details of the field and angle dependence.  Interestingly, we find that the pattern of the oscillations inverts at high temperature, similar to what was reported earlier for vortex-induced oscillations.
 
\section{Acknowledgements}
WAA acknowledges support by the Natural Sciences and Engineering Research Council (NSERC) of Canada. APK acknowledges support by the Deutsche Forschungsgemeinschaft (DFG, German Research Foundation)- project-ID-107745057-TRR 80.

\appendix

\section{Superconducting State}
\label{app:SC}
We consider the superconducting state of a single CuO$_2$ layer
with Rashba SOC.  The superconducting state is predominantly singlet,
with a $d$-wave symmetry, but additional triplet pieces are induced by
both the SOC and  the in-plane Zeeman field.  The goal of this
section is to evaluate the size of the triplet components.

Taking the basis
\begin{equation}
{\bf C}_\bk = \left[ \begin{array}{cccc} c_{\bk\uparrow}, & c_{\bk\downarrow}, & c^\dagger_{-\bk\uparrow}, & c^\dagger_{-\bk\downarrow} \end{array} \right ]^T,
\end{equation}
the Hamiltonian has the form ${\hat H} = {\sum_\bk}^\prime {\bf C}^\dagger_\bk {\bf H}_\bk {\bf C}_\bk$, where the primed sum is restricted to half of the Brillouin zone, and
\begin{equation}
{\bf H}_\bk = \left[ \begin{array}{cc} 
{\bf h}_\bk & {\bf \Delta}_\bk \\
{\bf \Delta}_\bk^\dagger & -{\bf h}^T_\bk 
\end{array} \right]
\label{eq:Hmat}
\end{equation}
with 
\begin{equation}
{\bf h}_\bk = \left[ \begin{array}{cc} 
\epsilon_\bk & \tilde g_\bk \\
\tilde g_\bk^\ast & \epsilon_\bk
\end{array} \right ]
\end{equation}
and
\begin{equation}
{\bf \Delta}_\bk = \left[ \begin{array}{cc} 
-d_{x\bk} + id_{y\bk} & d_{z\bk} + \chi_{d\bk} \\
d_{z\bk} - \chi_{d\bk} & d_{x\bk} + id_{y\bk}
\end{array} \right ].
\label{eq:DeltaK}
\end{equation}
where $\tilde g_\bk = \alpha (\sin k_y + i\sin k_x) -\mu B e^{-i\phi}$ and $\phi$ is the angle between the $x$-axis and the in-plane magnetic field.

We take the simplest form of pairing interaction appropriate for the cuprates,
\begin{eqnarray}
V(\bk-\bk') &=& V_0[ \cos (k_x-k_x') + \cos (k_y-k_y')] \nonumber \\
&=& V_0 \sum_{i} \eta_{i\bk} \eta_{i\bk'}
\end{eqnarray}
with 
\begin{eqnarray}
\eta_{d\bk} &=& \frac{1}{\sqrt{2}} (\cos k_x - \cos k_y), \\
\eta_{s\bk} &=& \frac{1}{\sqrt{2}} (\cos k_x + \cos k_y), \\
\eta_{x\bk} &=& \sin k_x, \\
\eta_{y\bk} &=& \sin k_y. 
\end{eqnarray}

Under the assumption that the singlet order parameter is $d$-wave, we write $\chi_{d\bk} = \chi_d \eta_{d\bk}$ with
\begin{eqnarray}
\chi_d &=&  \frac {V_0}4  \sum_{\bk'} \eta_{d\bk'}
\Big ( \langle c_{-\bk' \downarrow } c_{\bk' \uparrow } \rangle - \langle c_{-\bk' \uparrow } c_{\bk' \downarrow } \rangle \Big ). 
\end{eqnarray}
Similarly, the triplet components have odd spatial parity, and can therefore be written
\begin{equation}
d_{x\bk} = d_{xx} \eta_{x\bk} + d_{xy}\eta_{y\bk}, \mbox{ etc.} 
\end{equation}
with ($a=x,y$)
\begin{eqnarray}
d_{xa} &=& \frac {V_0}4  \sum_{\bk'} \eta_{a\bk'}
\Big ( \langle c_{-\bk' \downarrow } c_{\bk' \downarrow } \rangle - \langle c_{-\bk' \uparrow } c_{\bk' \uparrow } \rangle \Big ) \\
d_{ya} &=& \frac {V_0}{4i}  \sum_{\bk'} \eta_{a\bk'}
\Big ( \langle c_{-\bk' \downarrow } c_{\bk' \downarrow } \rangle + \langle c_{-\bk' \uparrow } c_{\bk' \uparrow } \rangle \Big ) \\
d_{za} &=&  \frac {V_0}4  \sum_{\bk'} \eta_{a\bk'}
\Big ( \langle c_{-\bk' \downarrow } c_{\bk' \uparrow } \rangle + \langle c_{-\bk' \uparrow } c_{\bk' \downarrow } \rangle \Big ). 
\end{eqnarray}
We choose parameters such that the singlet component of the order parameter is $\chi_d = 50$~meV, which is comparable to the antinodal gap in underdoped YBCO$_{6+x}$.

SOC induces a triplet component of the form $\pm d_{x\bk} + i d_{y\bk}$ that goes along with the singlet piece.\cite{Sigrist:2009gu}  For our model parameters, self-consistent calculations find that the triplet component is approximately 1\% of the dominant singlet component, and is nearly independent of magnetic field strength and direction.  In addition, the Zeeman field induces a second triplet component $d_{z\bk}$ that depends on field angle.  This component is found to be three orders of magnitude smaller than the singlet component.

\section{Thermal Conductivity: Large SOC Limit}
\label{sec:analytic}
In this appendix, we derive an analytic approximation for 
$\Pi^{xx}(x)$ that is valid in the limit $\alpha \gg \gamma, \mu B$.  To evaluate Eq.~(\ref{eq:Pi}), we transform both the Hamiltonian and the quasiparticle velocity operators to the helical basis.  When the SOC is large, we can neglect the mixing of bands of different helicities, either by impurities or by the magnetic field.  This simplification allows us to derive an explicit expression for $\Pi^{xx}(x)$.

\paragraph{The Hamiltonian in the helical basis.}
As a first step, we transform the Hamiltonian, Eq.~(\ref{eq:Hmat}), to
the helical basis via the unitary transformation,
\begin{equation}
{\cal U}_\bk = \left[ \begin{array}{cc} {\bf U}_\bk & 0 \\ 0 & {\bf U}_{-\bk}^\ast
\end{array} \right ]; \qquad
{\bf U}_\bk = \frac 1{\sqrt{2}}  \left[ \begin{array}{cc} 1 & - e^{i \theta_\bk} \\ e^{-i \theta_\bk} & 1\end{array}\right],
\end{equation}
with $e^{i  \theta_\bk} = { g_\bk}/{| g_\bk|}$ and $g_\bk = \alpha (\sin k_y + i \sin k_x)$.
In  zero field, ${\bf U}_\bk$ diagonalizes the nonsuperconducting Hamiltonian ${\bf
  h}_\bk$, with ${\bf U}^\dagger_\bk {\bf h}_\bk {\bf U}_\bk = \mbox{diag} (\xi_{\bk+}, \xi_{\bk-})$, and $\xi_{\bk \pm } = \epsilon_\bk \pm |g_\bk|$.  

In the superconducting state and with nonzero in-plane magnetic field, we obtain the transformed Hamiltonian,
\begin{eqnarray}
{\bf H}_\bk^\xi &=&  {\cal U}_\bk^\dagger {\bf H}_\bk {\cal U}_\bk 
=  \left [ \begin{array}{cc} 
{\bf h}^\xi_\bk &{\bf \Delta_\bk^\xi }\\
{\bf \Delta^\xi}^\dagger_\bk & - {\bf h}^\xi_{-\bk}
  \end{array}\right ].
\end{eqnarray}
The diagonal block is
\begin{equation}
  {\bf h}^\xi_\bk =
\left [ \begin{array}{cc}
\xi_{\bk+} -  \mu  B \cos(\theta_\bk  +\phi) &  i\mu B e^{i\theta_\bk}   \sin(\theta_\bk + \phi) \\
-i \mu B e^{-i\theta_\bk}   \sin(\theta_\bk + \phi) & \xi_{\bk-} + \mu  B \cos(\theta_\bk  +\phi) 
  \end{array} \right ]
\end{equation}
The off-diagonal
block has a similar structure to Eq.~(\ref{eq:DeltaK}); however, it
simplifies considerably if the the triplet components of the order parameter can be neglected.  Then
\begin{equation}
  \Delta^\xi_\bk = \left[ \begin{array}{cc} \Delta^+_\bk & 0 \\ 0 &
      \Delta^-_\bk \end{array}\right ],
\end{equation}
with $\Delta^\pm_\bk = -e^{\pm i\theta_\bk} \chi_{d\bk}$.  When $B=0$, the two helical bands are not mixed by singlet superconductivity under the restriction of zero-momentum pairing, and the Hamiltonian ${\bf H^\xi}_\bk$  thus describes two independent superconducting bands, each of which has a BCS-like structure.

\paragraph{The velocity matrix in the helical basis.}
Next, we transform the matrix defined by  Eq.~(\ref{eq:VSC}) via
\begin{equation}
{\bf V}^\xi_\bk = {\cal U}^\dagger_\bk {\bf V}_\bk {\cal U}_\bk.
\end{equation}
The top left block transforms as
\begin{eqnarray}
    {\bf U}_\bk^\dagger {\bf v}_\bk {\bf U}_\bk 
    &=& \left [ \begin{array}{cc} \nabla_\bk \xi^+_\bk & 0 \\ 0 & \nabla_\bk \xi^-_\bk \end{array} \right ] + i(\nabla_\bk \theta_\bk) \left [ \begin{array}{cc} 0 & g_\bk \\ -g_\bk^\ast & 0 \end{array} \right ], \nonumber \\
\end{eqnarray}
while the top right block is
\begin{eqnarray}
{\bf U}^\dagger_\bk {\bf v}_{\Delta, \bk} {\bf U}_{-\bk}^\ast &=& 
\left [ \begin{array}{cc}
-e^{i\theta_\bk} \nabla_\bk \chi_{d\bk} & 0 \\
0 & -e^{-i\theta_\bk} \nabla_\bk \chi_{d\bk}
\end{array} \right ].
\end{eqnarray}
Then
\begin{equation}
{\bf V}^\xi_\bk = \left [ \begin{array}{cccc}
{\bf v}^\xi_{\bk+} & {\bf v}^\mathrm{an}_\bk &  {\bf v}_{\Delta, \bk}^+ & 0  \\
({\bf v}^{\mathrm{an}}_\bk)^\ast &  {\bf v}^\xi_{\bk-} & 0 &  {\bf v}_{\Delta, \bk}^- \\
({\bf v}_{\Delta, \bk}^{+})^\ast & 0 & {\bf v}^\xi_{-\bk+} & ({\bf v}^\mathrm{an}_{-\bk})^\ast  \\
0 &  ({\bf v}_{\Delta, \bk}^-)^\ast & {\bf v}^\mathrm{an}_{-\bk} & {\bf v}^\xi_{-\bk-}  
\end{array} \right],
\end{equation}
where ${\bf v}^\xi_{\bk\alpha} = \nabla_\bk \xi_{\bk\alpha}$, ${\bf
  v}_{\Delta, \bk}^a = -e^{ia \theta_\bk} \nabla_\bk
\chi_{d\bk}$, and the SOC-related anomalous velocity is ${\bf v}^\mathrm{an}_\bk =
i(\nabla_k\theta_\bk) g_\bk$.

\paragraph{Re-ordering the basis.}
To move forward, it is useful to rearrange the Hamiltonian and velocity into blocks with given helicity, 
\begin{equation}
\left [ \begin{array} {c} \psi_{\bk+} \\ \psi_{\bk-}
    \\ \psi^\dagger_{-\bk+} \\ \psi^\dagger_{-\bk-} \end{array}\right
] \rightarrow \left [ \begin{array} {c} \psi_{\bk+}
    \\ \psi^\dagger_{-\bk+} \\ \psi_{\bk-}
    \\ \psi^\dagger_{-\bk-} \end{array}\right ].
\end{equation}
The velocity operator becomes
\begin{eqnarray}
{\bf V}^\xi_\bk &=& \left [ \begin{array}{cc|cc}
{\bf v}^\xi_{\bk+} &  {\bf v}_{\Delta, \bk}^+ &  {\bf v}^\mathrm{an}_\bk & 0  \\
({\bf v}_{\Delta, \bk}^{+})^\ast &  {\bf v}^\xi_{-\bk+} & 0 &  ({\bf v}^\mathrm{an}_{-\bk})^\ast  \\
\hline
({\bf v}^{\mathrm{an}}_\bk)^\ast &   0 & {\bf v}^\xi_{\bk-} &  {\bf v}_{\Delta, \bk}^- \\
0 &   {\bf v}^\mathrm{an}_{-\bk} & ({\bf v}_{\Delta, \bk}^-)^\ast &  {\bf v}^\xi_{-\bk-}  
\end{array} \right] \\
&=& {\bf V}^0_\bk + {\bf V}^\mathrm{an}_\bk
\label{eq:V0}
\end{eqnarray}
where ${\bf V}^0_\bk$ contains the diagonal blocks of ${\bf V}^\xi_\bk$ and ${\bf V}^\mathrm{an}_\bk$ contains the off-diagonal blocks.
The Hamiltonian is
\begin{equation}
  {\bf H}_\bk^\xi = \left [\begin{array}{cc} {\bf h}_{\bk+} & {\bf h}'_{\bk} \\ {\bf h}'_{\bk} & {\bf h}_{\bk -} \end{array}\right]
\end{equation}
with 
\begin{eqnarray}
{\bf h}_{\bk a} &=& \left [ \begin{array}{cc} \xi_{\bk a} -a \mu B
    \cos(\theta_\bk +\phi) &  \Delta^{a}_\bk \\    
    \Delta^{a \ast}_\bk &-\xi_{\bk a} - a \mu B \cos(\theta_\bk +\phi)
  \end{array} \right ] \nonumber \\
\end{eqnarray}
where $a = \pm$ labels the band helicity, and
\begin{eqnarray}
{\bf h}'_{\bk} &=&  i\mu B \sin(\theta_\bk + \phi) \left [ \begin{array}{cc}
    e^{-i\theta_\bk} & 0 \\
    0 & e^{i\theta_\bk} 
  \end{array} \right ].
\end{eqnarray}
The matrix ${\bf h}'_\bk$ mixes bands with different helicities and may be dropped when $\mu B \ll |\alpha|$.

\paragraph{Spectral function.}
 To zeroth order in ${\bf h}'_\bk$, ${\bf H}^\xi_\bk$ has the energy eigenvalues
\begin{eqnarray}
E_{\bk a s} &=& - a  \mu B \cos(\theta_\bk  +\phi) + sE_{\bk a}, \\
E_{\bk a} &=&  \sqrt{\xi_{\bk a}^2 + |\Delta_{\bk a}|^2} 
\end{eqnarray}
where $a=\pm$ is the helicity index and $s=\pm$ indicates whether the quasiparticle branch is upward or downward dispersing.
The Green's function is
\begin{eqnarray}
{\bf g}^0_{ab}(\bk,z) &=& 
\frac{\delta_{a,b}}{\tilde z_a^2  - E_{\bk a}^2} 
\left [\begin{array}{cc}
\tilde z_a+\xi_{\bk a}  & \Delta_{\bk a} \\
\Delta_{\bk a}^{\ast} &  \tilde z_a-\xi_{\bk a} 
\end{array}\right ] 
\end{eqnarray} 
where $a$ and $b$ represent the different helicities, and 
$\tilde z_\pm = z \pm \mu B \cos(\theta_\bk  +\phi)$ with $z$ a complex frequency.  To obtain the spectral function for (real) frequency $x$, we take
\begin{eqnarray}
{\bf a}^0_{ab}(\bk,x) &=& \frac{1}{2\pi i} \left [{\bf g}^0_{ab}(\bk,x-i\gamma) - {\bf g}^0_{ab}(\bk,x+i\gamma) \right ] \nonumber \\
&=& \frac{\delta_{a,b}}{2E_{\bk a}} \sum_\pm
\pm \delta(\pm E_{\bk a} - x - a  \mu B \cos(\theta_\bk  +\phi)) 
\nonumber \\
&&\times \left [\begin{array}{cc}
\xi_{\bk a} \pm E_{\bk a} & \Delta_{\bk a} \\
 \Delta_{\bk a}^\ast &  -\xi_{\bk a} \pm E_{\bk a} 
\end{array}\right ].
\label{eq:A0}
\end{eqnarray} 
In this expression, the $\delta$ functions are understood as Lorentzians, $\delta(x) = \pi^{-1} \gamma/(x^2+\gamma^2)$.

\paragraph{Thermal conductivity.}
We evaluate the transport kernel $\Pi(x)$ from Eq.~(\ref{eq:Pi}), using both the leading-order spectral function, Eq.~(\ref{eq:A0}), and the leading-order velocity operator ${\bf V}^0_\bk$  defined in Eq.~(\ref{eq:V0}), which neglects interband mixing due to the anomalous velocity term ${\bf v}^\mathrm{an}_\bk$. As noted before, this calculation is zeroth order in interband mixing, but not zeroth order in the magnetic field.   

\begin{figure}
	\includegraphics[width=0.5\textwidth]{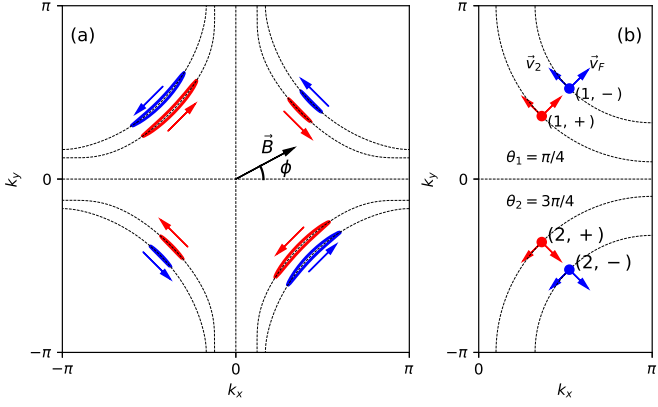}
	\caption{(a)  Bogoliubov Fermi surface pockets in the full Brillouin zone, along with the spin polarization of each quasiparticle branch near the node.  (b) Schematic of the reduced Brillouin zone, which consists of the region $0<k_x<\pi$, $-\pi < k_y < \pi$.  Nodal points, at which the superconducting $d$-wave order parameter vanishes, are indicated by dots. Arrows located at each dot indicate the directions of the normal and superconducting group velocities.   Local coordinate axes $k_1$ and $k_2$ are attached to each of the nodes, and are parallel to the velocities.  Each of the four nodes is labeled by its quadrant ($n=1,2$) and its helicity ($a=\pm$).    }
	\label{fig:bzone}
\end{figure}

We take the Brillouin zone to be $0<k_x <\pi$, $-\pi<k_y<\pi$, and sum over four nodal regions [Fig.~\ref{fig:bzone}(b)]; two of these correspond to a sum over helicty index $a$; the other two will be indicated by a sum over nodal index $n = 1,2$.  We use rotated momenta $k_1$ and $k_2$ near each node, with the understanding that they are rotated by $90^\circ$ between nodes in regions 1 and 2, and that the zero of the coordinate system is at the nodal point for each $(n,a)$.  The advantage of this definition is that we can write 
\begin{equation}
\xi_{\bk a} = \hbar v_F k_1; \qquad \Delta_{\bk a} = e^{ia\theta_{n}} \hbar v_2 k_2,
\end{equation}
for each node, with the approximation that $\theta_\bk$ can be treated as constant in the neighborhood of each node.  Recalling that $\tan \theta_\bk = \sin k_x/\sin k_y$, we obtain $\theta_1 = \pi/4$  for node 1 and $\theta_2 = 3\pi/4$ for node 2.  
Similarly, we will assume that the quasiparticle velocities depend only on the nodal index $n$, and are constant in the vicinity of each node.

Under these transformations,
\begin{eqnarray}
\frac{1}{N_k } {\sum_\bk}^\prime 
&\rightarrow& \frac{a_0^2}{2\pi^2} \int dk_x dk_y \\
&\rightarrow & 
\frac{a_0^2}{2\pi^2\hbar^2 v_Fv_2} \sum_{n=1}^2 \sum_{a=\pm} \int_0^\infty EdE \int_0^{2\pi} d\zeta, \nonumber \\ \label{eq:Jacobian}
\end{eqnarray}
where $a_0$ is the lattice constant.
The kernel for the thermal conductivity is therefore
\begin{eqnarray}
\Pi^{xx}(x) 
&=& \frac{1}{4\pi^2}  \sum_{a=\pm} \sum_{n=1}^2  \int_0^\infty \frac{EdE}{ v_F v_2} \oint d\zeta \nonumber \\
&& \times \mbox{Tr} \big [ {\bf a}_{n,aa}^0(E,\zeta,x)  {\bf v}^x_{n,aa}  {\bf a}_{n,aa}^0(E,\zeta,x) {\bf v}^x_{n,a a}  \big ], \nonumber \\
\label{eq:Pixx}
\end{eqnarray}
where we have made the linearized nodal approximation, $\xi = E\cos\zeta$, $\Delta_a = E e^{ia\theta_{n}} \sin \zeta$ and $\zeta \in [0,2\pi]$.  The trace in these equations is over particle-hole channels associated with superconductivity, so the matrices are $2\times 2$.
The velocity matrix ${\bf v}^x_{n,aa}$ is the $2\times 2$ matrix obtained from the top-left ($a=+$) or bottom-right ($a=-$) block of ${\bf V}^0_\bk$. The superscript $x$ refers to the component of the quasiparticle velocity along the $x$-direction.  Noting that $v^x_F = v_F/\sqrt{2}$ and $v^x_2 = -v_2/\sqrt{2}$ [Fig.~\ref{fig:bzone}(b)], 
\begin{equation}
{\bf v}^x_{n,aa} = \frac{1}{\sqrt{2}} \left [ \begin{array}{cc} v_F & - e^{ia\theta_n} v_2 \\ -e^{-ia\theta_n} v_2 & - v_F \end{array} \right ].
\end{equation}

In linearized nodal coordinates, the spectral function in Eq.~(\ref{eq:Pixx}) is
%
%
\begin{eqnarray}
{\bf a}^0_{n,aa}(E,\zeta,x) &=& \frac{1}{2} \sum_{s=\pm} 
 \delta\left [ sE - x - a\mu B \cos(\theta_n  +\phi) \right ] \nonumber \\
 && \times
\left [\begin{array}{cc}
1 + s\cos \zeta &  s e^{ia\theta_n} \sin \zeta \\
se^{-ia\theta_n} \sin \zeta &  1 - s\cos \zeta
\end{array}\right ] 
\end{eqnarray}
Separating the thermal conductivity kernel into contributions from the two helicity bands, $\Pi^{xx}(x) = \sum_a \Pi^a(x)$, we obtain for the $+$ helicity band
\begin{widetext}
\begin{eqnarray}
\Pi^+(x) &=& \frac{1}{4\pi^2 v_F v_2}  \left \{ \int_0^\infty dE\, E  \delta^2 \left [ E - x- \mu B \cos(\theta_1  +\phi) \right ] \right \} \nonumber \\
&&\times \frac 18
\oint d\zeta \mbox{Tr} \left [ 
\left [\begin{array}{cc} 1 + \cos \zeta &  e^{i\theta_1} \sin \zeta \\
e^{-i\theta_1} \sin \zeta &  1 - \cos \zeta \end{array}\right ]
\left [ \begin{array}{cc} v_F & - e^{i\theta_1} v_2 \\ -e^{-i\theta_1} v_2 & - v_F \end{array} \right ] 
\left [\begin{array}{cc} 1 + \cos \zeta &  e^{i\theta_1} \sin \zeta \\
e^{-i\theta_1} \sin \zeta &  1 - \cos \zeta \end{array}\right ]
\left [ \begin{array}{cc} v_F & - e^{i\theta_1} v_2 \\ -e^{-i\theta_1} v_2 & - v_F \end{array} \right ] \right ] \nonumber \\
&&+\frac{1}{4\pi^2 v_F v_2}  \left \{ \int_0^\infty dE\, E  \delta^2\left [ E + x + \mu B \cos(\theta_1  +\phi) \right ]  \right \} \nonumber \\
&&\times \frac 18
\oint d\zeta \mbox{Tr} \left [ 
\left [\begin{array}{cc} 1 - \cos \zeta &  -e^{i\theta_1} \sin \zeta \\
-e^{-i\theta_1} \sin \zeta &  1 + \cos \zeta \end{array}\right ]
\left [ \begin{array}{cc} v_F & - e^{i\theta_1} v_2 \\ -e^{-i\theta_1} v_2 & - v_F \end{array} \right ] 
\left [\begin{array}{cc} 1 - \cos \zeta &  -e^{i\theta_1} \sin \zeta \\
-e^{-i\theta_1} \sin \zeta &  1 + \cos \zeta \end{array}\right ]
\left [ \begin{array}{cc} v_F & - e^{i\theta_1} v_2 \\ -e^{-i\theta_1} v_2 & - v_F \end{array} \right ]    \right ] \nonumber \\
&&+\frac{2}{4\pi^2 v_F v_2}   \left \{ \int_0^\infty dE\, E  \delta\left [ E -x- \mu B \cos(\theta_1  +\phi) \right ] \delta\left [ E +x+ \mu B \cos(\theta_1  +\phi) \right ]  \right \} \nonumber \\
&&\times \frac 18
\oint d\zeta \mbox{Tr} \left [ 
\left [\begin{array}{cc} 1 + \cos \zeta &  e^{i\theta_1} \sin \zeta \\
e^{-i\theta_1} \sin \zeta &  1 - \cos \zeta \end{array}\right ]
\left [ \begin{array}{cc} v_F & - e^{i\theta_1} v_2 \\ -e^{-i\theta_1} v_2 & - v_F \end{array} \right ] 
\left [\begin{array}{cc} 1 - \cos \zeta &  -e^{i\theta_1} \sin \zeta \\
-e^{-i\theta_1} \sin \zeta &  1 + \cos \zeta \end{array}\right ]
\left [ \begin{array}{cc} v_F &  -e^{i\theta_1} v_2 \\ -e^{-i\theta_1} v_2 & - v_F \end{array} \right ]  \right ]. \nonumber \\
&&+ \left( \theta_1 \rightarrow \theta_2 \right )
\end{eqnarray}
Using the Lorentzian form for the delta-function, the energy integrals easily give
\begin{eqnarray}
\frac{\gamma^2}{\pi^2} \int_0^\infty dE\, \frac{E}{[(E \mp \tilde x )^2+\gamma^2]^2} &=& \frac{1}{2\pi^2} \left [
1 \pm \frac{\tilde x}{\gamma} \left( \frac {\pi}2 \pm \tan^{-1} \frac {\tilde x} \gamma \right ) 
\right ], 
\label{eq:E-integral} \\
%
\frac{\gamma^2}{\pi^2} \int_0^\infty dE\, \frac{E}{[(E-\tilde x)^2+\gamma^2][ (E+\tilde x)^2+\gamma^2] } &=& \frac{1}{2\pi^2}  \frac{\gamma}{\tilde x}  \tan^{-1} \frac {\tilde x}\gamma. 
\end{eqnarray}
Performing the matrix multiplications, taking the traces, and integrating over the angle $\zeta$ gives
\begin{eqnarray}
\Pi^+(x) &=& \frac {(v_F^2+v_2^2)}{8\pi^3 v_Fv_2} \Bigg \{ 
2 + {\cal E}\left[ \frac{x + \mu B \cos(\frac \pi 4 + \phi)}{\gamma} \right ]
+ {\cal E}\left[ \frac{x + \mu B \cos(\frac {3\pi} 4 + \phi)}{\gamma} \right ] \Bigg \},  
\label{eq:Piplus}
\end{eqnarray}
where
\begin{equation}
{\cal E}(x) = \left( x + \frac 1x \right ) \tan^{-1} x.
\end{equation}

A nearly identical calculation gives the contribution $\Pi^-(x)$ to the thermal conductivity kernel from the $-$ helicity bands.  $\Pi^-(x)$ has teh same form as Eq.~(\ref{eq:Piplus}), but with $\mu B \rightarrow -\mu B$.  The total kernel is then
\begin{eqnarray}
\Pi^{xx}(x) &=& \frac{1}{8\pi^3}\frac {v_F^2+v_2^2}{ v_Fv_2} \Bigg \{ 
4 + \sum_\pm {\cal E}\left[ \frac{x \pm \mu B \cos(\frac \pi 4 + \phi)}{\gamma} \right ]
+  \sum_\pm {\cal E}\left[ \frac{x \pm \mu B \cos( \frac{3\pi}{4} + \phi)}{\gamma} \right ] \Bigg \}.
\end{eqnarray}

\end{widetext}

The thermal conductivity follows from
\begin{eqnarray}
\frac{\kappa_{xx}}{T} = \frac{ \pi}{\hbar d T^2} \int dx\, x^2 \frac {\partial f(x)}{\partial x} \Pi^{xx}(x)
\end{eqnarray}
In the limit $T\rightarrow 0$, this expression simplifies to
\begin{eqnarray}
\frac{\kappa_{xx}}{T} &\rightarrow&  \frac{ k_B^2 \pi^3}{3 \hbar d} \Pi^{xx}(0) \nonumber \\
&=& \frac {k_B^2}{12 \hbar d} \frac{v_F^2+v_2^2}{v_F v_2}
\Bigg \{ 
2 +  {\cal E}\left[ \frac{\mu B \cos(\frac \pi 4 + \phi)}{\gamma} \right ] \nonumber \\
&&  +  {\cal E}\left[ \frac{\mu B \cos( \frac{3\pi}{4} + \phi)}{\gamma} \right ] \Bigg \}.
\end{eqnarray}
As an important check, notice that as $\mu B \rightarrow 0$, ${\cal E}\rightarrow 1$, and 
\begin{equation}
\frac{\kappa_{xx}^{0}}{T} \rightarrow  \frac{k_B^2} {3 \hbar d } \frac{v_F^2  + v_2^2}{v_F v_2} 
\end{equation}
which is the result first worked out by Durst and Lee.\cite{Durst:2000iw}


\section{Density of States: Large SOC Limit}
\label{sec:DOS}
In this section, we derive an expression for the density of states induced by the in-plane magnetic field in the limit of large SOC, namely $|\alpha| \gg \mu B, \, \gamma$.

From Eq.~(\ref{eq:A0}), the spectral function is approximately
\begin{eqnarray}
{\bf a}^0_{aa} (\bk, x) &=& \frac 12 \sum_{s=\pm 1}
\delta [sE_{\bk a} - x -a \mu B \cos(\theta_\bk +\phi) ]
\nonumber \\ && \times
\left [ \begin{array}{cc}
1 +s \cos \zeta & s \sin \zeta \\
s \sin \zeta & 1-s \cos \zeta
\end{array}	\right ].
\end{eqnarray}
For the residual density of states at the Fermi energy, we set $x=0$ and sum $\bk$ over the reduced Brillouin zone [Fig.~\ref{fig:bzone}(b)].   Since both helicities make identical contributions to the density of states, we calculate the result for the positive helicity and multiply the result by 2.  Making use of Eq.~(\ref{eq:Jacobian}), we obtain
\begin{eqnarray}
\rho(\varepsilon_F) &=& \frac{2}{N_k}  {\sum_\bk}' [{\bf a}^0_{++}(\bk, 0)]_{11} \nonumber \\
&=& \frac{a_0^2}{2\pi^2v_Fv_2} \sum_{n=1}^2 \sum_{s=\pm} \oint d\zeta \int_0^\infty EdE 
\nonumber \\ && \times
\delta[ sE + \mu B\cos (\theta_n+\phi) ]  
\end{eqnarray}
where $\theta_1 = \frac \pi 4$ and $\theta_2 = \frac {3\pi}{4}$.
Again, substituting Lorentzians for the delta-functions 
we get
\begin{widetext}
\begin{eqnarray}
\rho(\varepsilon_F) &=& \frac{a_0^2\gamma}{\pi^2v_Fv_2}\sum_{n=1}^2 \left [ \ln \frac{\Lambda^2}{[\mu B\cos(\theta_n+\phi)]^2 + \gamma^2} + 2\frac{\mu B\cos (\theta_n+\phi)}{\gamma} \tan^{-1} \frac{\mu B\cos (\theta_n+\phi)}{\gamma} \right ],
\end{eqnarray}
where $\Lambda$ is a cutoff.  The change in the DOS induced by the magnetic field is obtained by subtracting off the $B=0$ result,
\begin{eqnarray}
\Delta \rho(\varepsilon_F) &=& \frac{2a_0^2}{\pi^2v_Fv_2}\sum_{n=1}^2 \left [\gamma \ln \frac{\gamma}{\sqrt{[\mu B\cos(\theta_n+\phi)]^2 + \gamma^2}} + \mu B\cos (\theta_n+\phi) \tan^{-1} \frac{\mu B\cos (\theta_n+\phi)}{\gamma} \right ]
\label{eq:drho}
\end{eqnarray}
\end{widetext}
The change in the linear specific heat coefficient due to the magnetic field is then
\begin{equation}
\Delta \gamma_0 = \lim_{T\rightarrow 0} \frac {\Delta c_v}{T} = \frac{\pi^2}{3} k_B^2 \Delta \rho(\epsilon_F).
\end{equation}


\bibliography{Cuprate_SOC}

\end{document}